\pdfoutput=1

\RequirePackage{fix-cm}
\documentclass{svjour3}                     
\smartqed  
\usepackage{graphicx}
\usepackage{epstopdf}
\usepackage{framed}

\usepackage{enumitem}
\setitemize[1]{itemsep=0pt,partopsep=0pt,parsep=\parskip,topsep=0pt}

\usepackage{framed}  
\usepackage{booktabs}
\usepackage[misc]{ifsym}
\usepackage{mathptmx}      
\usepackage{latexsym}
\usepackage{hyperref}
\usepackage{amsmath}
\hypersetup{hypertex=true,
	colorlinks=true,
	linkcolor=blue,
	anchorcolor=blue,
	citecolor=blue}
\begin{document}
	
	\title{Quantum k-means algorithm based on Trusted server in Quantum Cloud Computing
	}

	
	\author{Changqing Gong\textsuperscript{1}   \and
		Zhaoyang Dong\textsuperscript{1} \and
		Abdullah Gani\textsuperscript{2} \and
		Han Qi\textsuperscript{1} 
	}
	\institute{
		\Letter \quad Han Qi \at 
		qihan@sau.edu.cn
		\and
		\textsuperscript{1} \quad School of Computer Science and Technology, Shenyang Aerospace University, Shenyang, China.\\	\textsuperscript{2} \quad Faculty of computer science and information technology, University of Malaya, Malaysia.
		\at
	}
	
	\date{Received: date / Accepted: date}

	\maketitle
	
	\begin{abstract}
		We propose a quantum k-means algorithm based on quantum cloud computing that effectively solves the problem that the client can not afford to execute the same quantum subroutine repeatedly in the face of large training samples. In the quantum k-means algorithm, the core subroutine is the Quantum minimization algorithm ($GroverOptim$), the client needs to repeat several Grover searches to find the minimum value in each iteration to find a new clustering center, so we use quantum homomorphic encryption scheme (QHE) to encrypt the data and upload it to the cloud for computing. After calculation, the server returns the calculation result to the client. The client uses the key to decrypt to get the plaintext result. It reduces the computing pressure for the client to repeat the same operation. In addition, when executing in the cloud, the key update of T-gate in the server is inevitable and complex. Therefore, this paper also proposes a $T$-gate update scheme based on trusted server in quantum ciphertext environment. In this scheme, the server is divided into trusted server and semi-trusted server. The semi-trusted server completes the calculation operation, and when the T-gate is executed in the circuit, the trusted server assists the semi-trusted server to calculate the T-gate, and then randomly generates a key and uploads it to the semi-trusted server. The trusted server assists the client to complete the key update operation, which once again reduces the pressure on the client and improves the efficiency of the quantum homomorphic encryption scheme. And on the basis of this scheme, the experiment of calculating the similarity between ciphertext quantum states ($SwapTest$) and retrieving the minimum ciphertext ($GroverOptim$) is given by using IBM Qiskit to give the subroutine of quantum k-means. The experimental results show that the scheme can realize the corresponding computing function on the premise of ensuring security.
	\end{abstract}
	
	\keywords{Quantum computing, Quantum homomorphic encryption, Key update algorithm, IBM Quantum Qiskit, Quantum cloud computing}
	
	\section{Introduction}
	\label{intro}
	
	Both quantum computing and cloud computing are technologies that can change the way computing in the future. Quantum computing uses the coherence and entanglement properties of quantum physics to create some high-speed computing models and accelerate classical algorithms. Cloud computing can provide computing power as a service to clients. The combination of the two to realize a new kind of quantum cloud computing will be the focus of future computing services.\cite{devitt2016performing(1)}
	
	In recent years, with the gradual development of quantum computing, some very important quantum algorithms have been proposed. e.g. Shor\cite{shor1994algorithms(2)} proposed a quantum algorithm for finding discrete logarithms and prime factorization. In 1996, Grover's quantum search algorithm\cite{grover1996fast(3)} proposed to solve the search problem in disordered data database, and this algorithm is faster than all known search algorithms. Aïmeur et al\cite{aimeur2006machine(4)} proposed a quantum subroutine $SwapTest$ which uses probability to reveal the similarity of two quantum states. Based on the proposals of these algorithms, some scholars have widely used them as subroutines in various quantum machine learning algorithms. Anguita\cite{anguita2003quantum(5)} used the Grover quantum search algorithm to improve the training efficiency in the support vector machine (SVM) to achieve the effect of optimizing the SVM. Ruan et al.\cite{ruan2014quantum(6)} proposed a quantum principal component analysis algorithm QPCA applied to face recognition, and used quantum states to encode facial features, and used Grover quantum search algorithm to achieve a secondary acceleration effect on face recognition. In Ref\cite{PhysRevLett.113.130503(7)},Rebentrost et al. proposed a quantum support vector machine QSVM, which uses quantum computing to solve the inner product calculation of training data, that is, uses quantum computing to solve the deviation of the matrix to obtain the normalized kernel matrix of the training data. Lu et al.\cite{lu2014quantum(8)} proposed a quantum decision tree algorithm, which uses the SwapTest subroutine to calculate the fidelity between quantum states instead of the similarity between training data, according to which the training set is divided into subclasses, and quantum information entropy is introduced as the basis for selecting decision nodes to establish a quantum decision tree. Durr and Hoyer proposed the quantum minimum algorithm\cite{durr1996quantum(9)}, which is widely used in various quantum machine learning algorithms as an extension of the Grover algorithm as known as $GroverOptim$. The algorithm uses the Quantum Oracles in the Grover search to mark those elements below an arbitrary threshold until it converges to the global minimum.
	
	As a classic clustering algorithm, k-means has been widely used in image recognition, medical health, data prediction, etc. since its proposal\cite{2010Application(10),xing2003distance(11),liu2020semi(12)}. However, in the era of big data, through the strong parallel computing power of quantum computing, Aïmeur\cite{aimeur2013quantum(13)} proposed the quantum k-means algorithm, which uses $SwapTest$ to calculate the similarity between the training vector and the cluster center, and $GroverOptim$ assigns the vector to the nearest The centroid of the cluster. But when the client does not have strong quantum computing capabilities, repeated calculation of $GroverOptim$ will still put a lot of pressure on the client.
	
	On the other hand, quantum computing will be realized in the near future, providing quantum computing services to clients through quantum cloud. In order to ensure the security of quantum cloud computing, when Rohde et al.\cite{rohde2012quantum(14)} studied the quantum walk on encrypted data in 2012, they proposed a restricted symmetric homomorphic encryption scheme based on the boson sampling model and the quantum walk model. In 2013, Liang\cite{liang2013symmetric(15)} first proposed quantum homomorphic encryption (QHE) based on Quantum one-time pad (QOTP)\cite{boykin2003optimal(16)}, which is mainly composed of key generation algorithm, encryption algorithm, evaluation algorithm and decryption algorithm. While researchers propose a quantum homomorphic encryption scheme, they also pay attention to the research of key update algorithm. As early as 2005, there was no concept of quantum homomorphic encryption, Ref\cite{childs2001secure(17)} gave the key update algorithm of related quantum gates when studying security-assisted quantum computing, and considered its security in detail. When there is a T gate in the circuit, the solution requires a two-way quantum interaction process, and the client needs to have the ability to execute a quantum swap gate. In 2014, Fisher et al.\cite{fisher2014quantum(18)} gave a key update algorithm for quantum computing on encrypted data. In this scheme, when the T gate is to be executed, the scheme no longer requires two-way quantum interaction, but one-way Quantum interaction and two-way classical interaction. In addition, Fisher et al. also made experiments on real quantum devices of the scheme, verifying the correctness of the scheme. In 2015, Liang gave a quantum fully homomorphic encryption scheme based on general-purpose circuits by studying general-purpose quantum circuits\cite{liang2015quantum(19)}, which is similar to Fisher et al.'s scheme, except that the key update algorithm of this scheme only relies on general-purpose quantum circuits. The structure of the circuit has nothing to do with the algorithm of the server. but in Fisher's solution, the client needs to know the operation done by the server to complete the key update. In the subsequent research,  Liang et al. construct a symmetric quantum homomorphic encryption scheme and an asymmetric quantum full-homomorphic encryption scheme based on quantum fault-tolerant\cite{liang2015quantum(20)}. In the symmetric quantum homomorphic encryption scheme, the private key is a quantum error-correcting code, and the client needs to provide some auxiliary qubits. In the asymmetric quantum homomorphic encryption scheme, there is a periodic interaction between the client and the server. In 2020, Liang\cite{liang2020teleportation(21)} proposed a quantum homomorphism encryption scheme based on gate teleportation by studying the teleportation of quantum gates and combining the quantum one-time pad scheme (QOTP). In this scheme, the key update process of T-gate does not need interactive process, and it is perfectly secure. 
	
	Recently, some scholars have gradually paid attention to the problem of ciphertext computing in quantum cloud computing. In 2017, a quantum homomorphic encryption experiment was proposed based on IBM cloud quantum computing platform\cite{Huang2017Homomorphic}. The experiment encrypts the input of the matrix inversion algorithm (HHL algorithm) by introducing a key, and decrypts the results after cloud computing. However, the process does not use any homomorphic encryption scheme, and its security can not be guaranteed. In the same year, Sun et al.\cite{sun2017efficient(22)} proposed a symmetric quantum partial homomorphic encryption scheme , in which the evaluation function is independent of the key. Based on this quantum homomorphic encryption scheme, an effective symmetric searchable encryption scheme is given and proved to be secure. However, the search algorithm given in this scheme is linear search, and the efficiency will be very low when the search space is very large. In 2020, Zhou et al.\cite{zhou2020quantum(23)} proposed a search scheme on encrypted data based on quantum homomorphic encryption. In this scheme, Grover algorithm is applied to ciphertext data, which realizes parallel search and improves the retrieval efficiency. However, when there is a T-gate in the circuit, the process is tedious. 
	
	So far, quantum cloud computing based on quantum homomorphism are mainly  ciphertext retrieval schemes, but other computing for ciphertext are very rare. Therefore, in order to improve the shortcomings of the current research situation, this paper proposes a quantum k-means algorithm based on trusted server in quantum cloud.
	
	\begin{itemize}
		\item[$ \bullet $] In order to solve the problem that T-gate update is too tedious in the current quantum homomorphic encryption scheme\cite{liang2020teleportation(21)}, a $T$-gate update scheme based on trusted server is proposed. In this scheme, a trusted quantum trusted server is introduced to assist the semi-trusted server to update the T-gate. At the same time, the trusted server also needs to help the client complete the key update operation to reduce the pressure on the client again.
		\item[$ \bullet $] Combined with the improved quantum homomorphic encryption scheme, the quantum k-means algorithm is improved, and the core subprograms that need to be executed repeatedly are executed in the quantum cloud. When the amount of data is large, the client only needs to measure the results, and uses the decryption key to decrypt the returned results, which does not need additional quantum computing power, which reduces the pressure on the client. The efficiency of the key updata algorithm is improved.
		\item[$ \bullet $] We prove the quantum subroutines $SwapTest$ and $GroverOptim$ on ciphertext based on IBM Qiskit. The experimental results show that the improved quantum k-means algorithm based on trusted server is feasible, but there are noise and imperfect gates in the experimental environment, which will cause errors to the experimental results.
	\end{itemize}
	\vbox{}
		
	The rest of this article is organized as follows. We summarize preliminaries knowledge of quantum computing in Sect.\ref{Preliminaries}. In the section\ref{3}, we propose a T-gate update scheme based on trusted server. In Sect.\ref{4}, using the improved quantum homomorphism encryption scheme, a quantum K-means algorithm based on quantum cloud computing is proposed. Experimental verification of the quantum core subroutines executed in the quantum cloud server based on IBM Qiskit in Sect.\ref{5} and\ref{6}. In Sect.\ref{7}, the proposed scheme is analyzed in terms of safety and time efficiency. Finally, the conclusion and future work are given in Sect.\ref{8}.
	
	\section{Preliminaries}
	\label{Preliminaries}
	\subsection{Quantum gate}
\begin{figure}
		\includegraphics[width=0.8\linewidth]{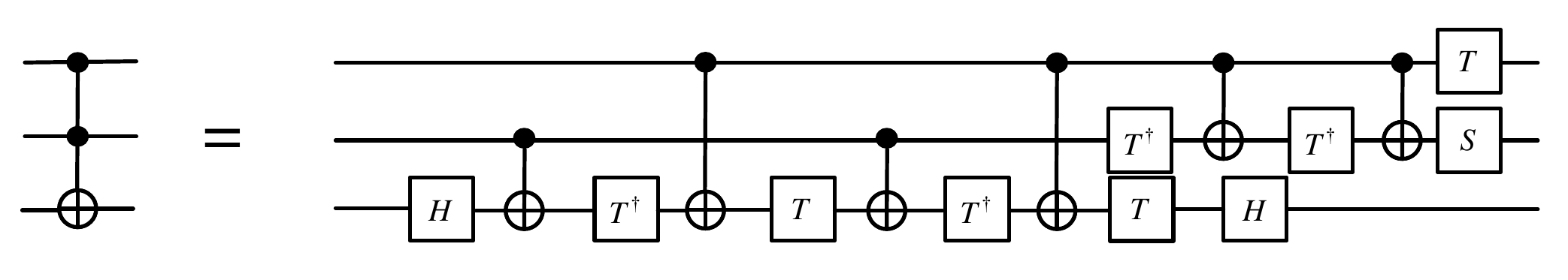}
		\caption{Quantum circuit representation of Toffoli gate and its combined realization.}
		\label{figure1}
	\end{figure}

	The basic unit of the quantum computing model is the qubit gate. The specific mathematical form of the basic single qubit gate $G = \left\{ {X,Y,Z,H,S,T} \right\}$ can refer to Ref\cite{nielsen2002quantum(24)}. This section introduces some complex multi-qubit gates used in this article. First, the two-qubit controlled NOT gate (CNOT or written as CX), switch gate (SWAP), and controlled Z gate (CZ) can be expressed as:\[CNOT = \left[ {\begin{array}{*{20}{c}}
		1&0&0&0\\
		0&1&0&0\\
		0&0&0&1\\
		0&0&1&0
		\end{array}} \right]\quad SWAP = \left[ {\begin{array}{*{20}{c}}
		1&0&0&0\\
		0&0&1&0\\
		1&0&0&0\\
		0&0&1&0
		\end{array}} \right]\quad CZ = \left[ {\begin{array}{*{20}{c}}
		1&0&0&0\\
		0&1&0&0\\
		0&0&1&0\\
		0&0&0&{ - 1}
		\end{array}} \right]\]
	The three-qubit controlled-controlled NOT gate (Toffoli) is also a quantum gate commonly used in quantum computing. It can be implemented by a combination of H gate, S gate, CNOT gate and T gate. The circuit diagram is shown in the Fig.\ref{figure1}. The Toffoli gate includes three qubits, two control qubits and one target qubit. Its function is to flip the target qubit when the two control qubits are $\left| 1 \right\rangle $ at the same time. 
	
\begin{figure}
		\includegraphics[width=0.6\linewidth]{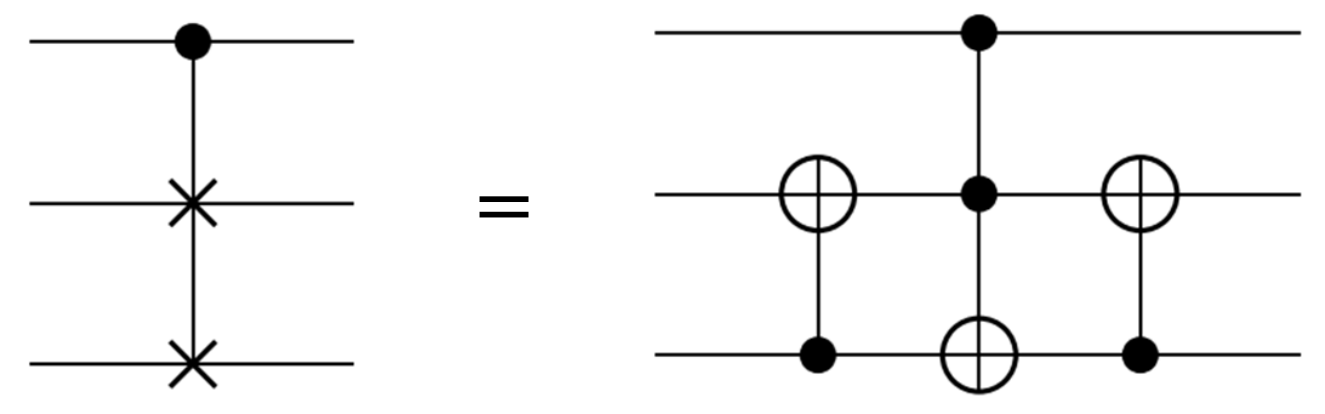}
		\caption{Quantum Circuit representation of C-SWAP Gate and its Combinatorial implementation.}
		\label{figure2}
	\end{figure}

	The quantum circuit of the other three qubit gate controlled swap gate (C-SWAP) is shown in Fig.\ref{figure2}. The C-SWAP gate also requires three qubits, of which the first register is the control bit and the other two are controlled bits. Its function is to exchange the quantum state between the second register and the third register when the first register is $\left| 1 \right\rangle $.
	
	\subsection{Quantum key update algorithm based on Quantum One-time Pad (QOTP)}
	
	Boykin and Roychowdhury proposed Quantum one-time pad algorithm (QOTP)\cite{boykin2003optimal(16)}, in this scheme, the client uses the combination of Pauli X and pauli Z to encrypt the plaintext $\left| \varphi  \right\rangle $ to get the ciphertext ${X^a}{Z^b}\left| \varphi  \right\rangle $, where $a,b$ is randomly selected in $\left\{ {0,1} \right\}$.
	
	For encrypted $n $qubit data, the encryption key is the $ek = \left( {{a_0},{b_0}} \right)$ and decryption key $dk = \left( {{a_m},{b_m}} \right)$. Quantum computation $U$ can be regarded as composed of quantum gate $O \in \left\{ {H,I,X,Y,Z,S,CNOT,T,{T^\dag }} \right\}$. We define that when the $N$ gate is applied to the $m$-th qubit (or the  $m$-th qubit and $l$-th qubit), the $p$-th key update can be expressed as
	\begin{equation}
	G\left[ N \right]{X^{{a_p}}}\left( m \right){Z^{{b_p}}}\left( m \right)\left| \varphi  \right\rangle {\rm{ = }}{X^{{a_{p + 1}}}}\left( m \right){Z^{{b_{p + 1}}}}\left( m \right)G\left[ N \right]\left| \varphi  \right\rangle 
	\end{equation}
	According to the relationship of the encryption door used in encryption, the following key update rules are established:
	\begin{itemize}
		\item[$ \bullet $] If $G\left[ N \right]$ is not applied to the qubits of the quantum circuit, then $\left( {{a_{p{\rm{ + }}1}}\left( m \right),{b_{p{\rm{ + }}1}}\left( m \right)} \right) = \left( {{a_p}\left( m \right),{b_p}\left( m \right)} \right)$.
		\item[$ \bullet $] If $G\left[ N \right] \in \left\{ {{H_m}} \right\}$, then $\left( {{a_{p{\rm{ + }}1}}\left( m \right),{b_{p{\rm{ + }}1}}\left( m \right)} \right) = \left( {{b_p}\left( m \right),{a_p}\left( m \right)} \right)$.
		\item[$ \bullet $] If $G\left[ N \right] \in \left\{ {{I_m},{X_m},{Y_m},{Z_m}} \right\}$,then $\left( {{a_{p{\rm{ + }}1}}\left( m \right),{b_{p{\rm{ + }}1}}\left( m \right)} \right) = \left( {{a_p}\left( m \right),{b_p}\left( m \right)} \right)$.
		\item[$ \bullet $] If $G\left[ N \right] \in \left\{ {{S_m}} \right\}$, then $\left( {{a_{p{\rm{ + }}1}}\left( m \right),{b_{p{\rm{ + }}1}}\left( m \right)} \right) = \left( {{a_p}\left( m \right),{a_p}\left( m \right) \oplus {b_p}\left( m \right)} \right)$.
		\item[$ \bullet $] If $G\left[ N \right] \in \left\{ {CNO{T_{m,l}}} \right\}$, then $\begin{array}{l}
		\left( {{a_{p{\rm{ + }}1}}\left( m \right),{b_{p{\rm{ + 1}}}}\left( m \right)} \right) = \left( {{a_p}\left( m \right),{b_p}\left( m \right) \oplus {b_p}\left( l \right)} \right)\\
		\;\;\,\left( {{a_{p{\rm{ + }}1}}\left( l \right),{b_{p{\rm{ + }}1}}\left( l \right)} \right) = \left( {{a_p}\left( m \right),{a_p}\left( l \right) \oplus {b_p}\left( l \right)} \right)
		\end{array}$.
		\item[$ \bullet $] If $G\left[ N \right] \in \left\{ {{T_m},{T^\dag }_m} \right\}$, At this point, the client needs to perform a rotation measurement $\Phi \left( {{S^{{a_{p - 1}}\left( m \right)}}} \right)$ on $G\left[ N \right]$, and the measurement result is $\left( {{r_a}\left( m \right),{r_b}\left( m \right)} \right)$. The key update rule at this time is as follows:	
		
		\item When $G\left[ N \right] = {T_m}$, then $\left( {{a_{p{\rm{ + }}1}}\left( m \right),{b_{p{\rm{ + }}1}}\left( m \right)} \right) = \left( {{a_p}\left( m \right) \oplus {r_a}\left( l \right),{a_{p - 1}}\left( m \right) \oplus {b_p}\left( m \right) \oplus {r_b}\left( l \right)} \right)$.
		
		\item When $G\left[ N \right] = T_m^\dag $, then $\left( {{a_{p{\rm{ + }}1}}\left( m \right),{b_{p{\rm{ + }}1}}\left( m \right)} \right) = \left( {{a_p}\left( m \right) \oplus {r_a}\left( l \right),{b_p}\left( m \right) \oplus {r_b}\left( l \right)} \right)$.
		
		In addition, through the corresponding relationship between $CZ$ gate and Pauli $X$ and $Z$ gate, we derive the key update formula corresponding to $CZ$ gate. The key update rule is as follows:
		\item[$ \bullet $] If $G\left[ N \right] \in \left\{ {C{Z_{m,l}}} \right\}$, then $\begin{array}{l}
		\left( {{a_{p{\rm{ + }}1}}\left( m \right),{b_{p{\rm{ + 1}}}}\left( m \right)} \right) = \left( {{a_p}\left( m \right),{b_p}\left( m \right) \otimes {a_p}\left( l \right)} \right)\\
		\;\;\,\left( {{a_{p{\rm{ + }}1}}\left( l \right),{b_{p{\rm{ + }}1}}\left( l \right)} \right) = \left( {{a_p}\left( l \right),{b_p}\left( l \right) \otimes {a_p}\left( m \right)} \right)
		\end{array}$.
	\end{itemize}
	
	Quantum gate: I gate, X gate, Z gate, H gate and S gate, and double qubit CNOT gate all belong to Clifford gate. When the server needs to execute Clifford gate on ciphertext data, it does not need any additional resources, only the transformation of Pauli matrix X and Z. When non-Clifford gates act on encrypted data, such as $T$ and ${T^\dag }$ gates, $S$ gates are used. i.e.$T{X^a}{Z^b}\left| \varphi  \right\rangle  = {X^a}{Z^{a \oplus b}}{S^a}T\left| \varphi  \right\rangle $. Therefore, in order to remove the S-gate error caused by T-gate key update, this paper presents a new T-gate processing method based on trusted server-assisted scheme. Compared with the previous work, this scheme is more clear to deal with T-gate, and it is suitable for the case of high $T$-gate complexity. 
	
	\section{$T$-gate update scheme based on trusted server in quantum cloud}
	\label{3}
\begin{figure}
		\includegraphics[width=0.6\linewidth]{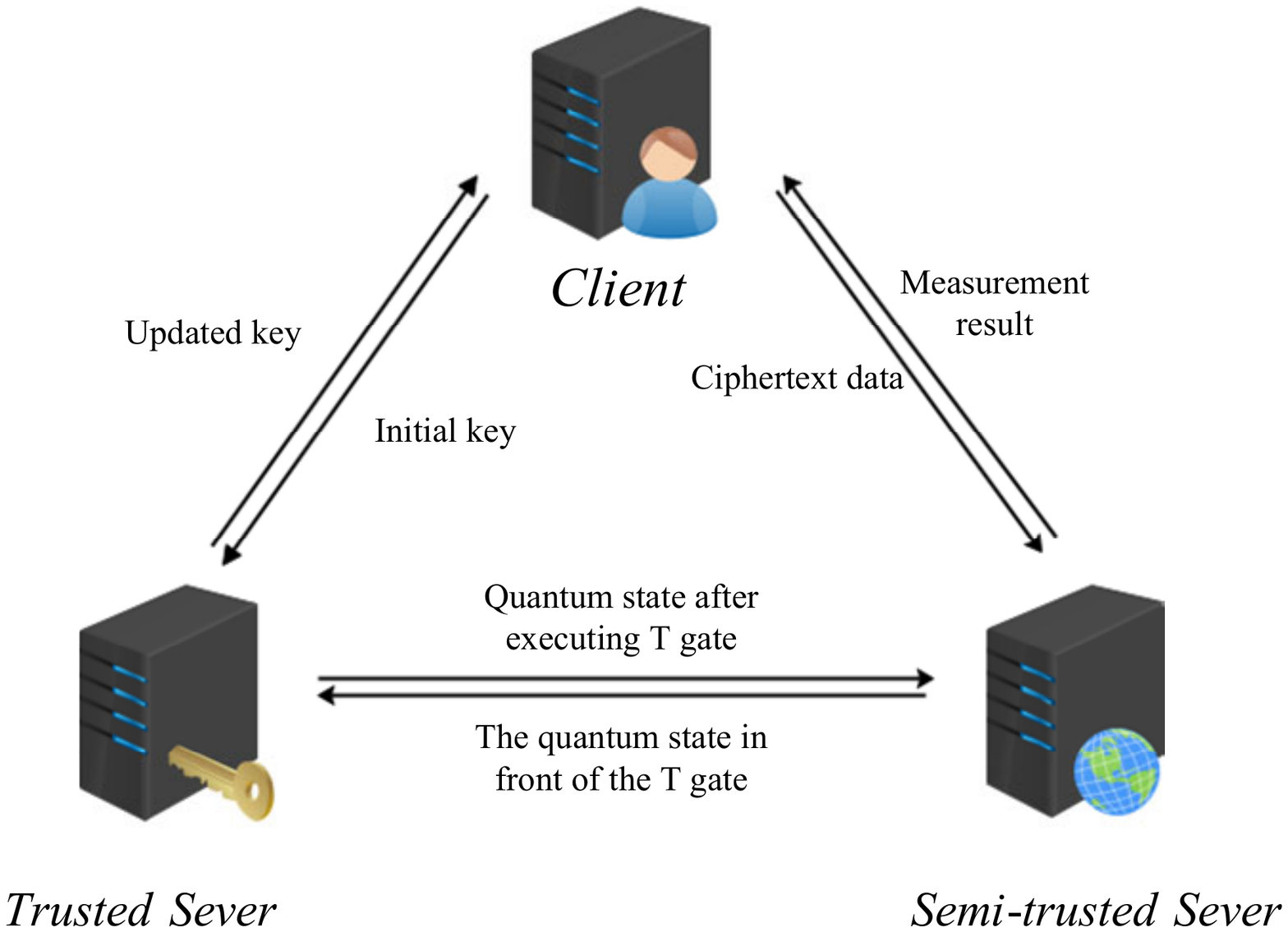}
		\caption{$T$-gate update process based on trusted server.}
		\label{figure3}
	\end{figure}

	In the quantum homomorphic encryption scheme (QHE)\cite{liang2020teleportation(21)}, the complexity of the update process of $T$-gates is based on the number of $T$-gates, the client needs to prepare the same number of bell states and measure them. When the complexity of the $T$-gate in the quantum circuit is high, the operation that needs to be performed will take more time. In order to solve this problem, we propose a T-gate update scheme based on trusted server. Our scheme architecture is divided into three parts, client, semi-trusted server and trusted server. The framework of the update scheme is shown in the Fig.\ref{figure3}.
	
\begin{figure}
		\includegraphics[width=0.9\linewidth]{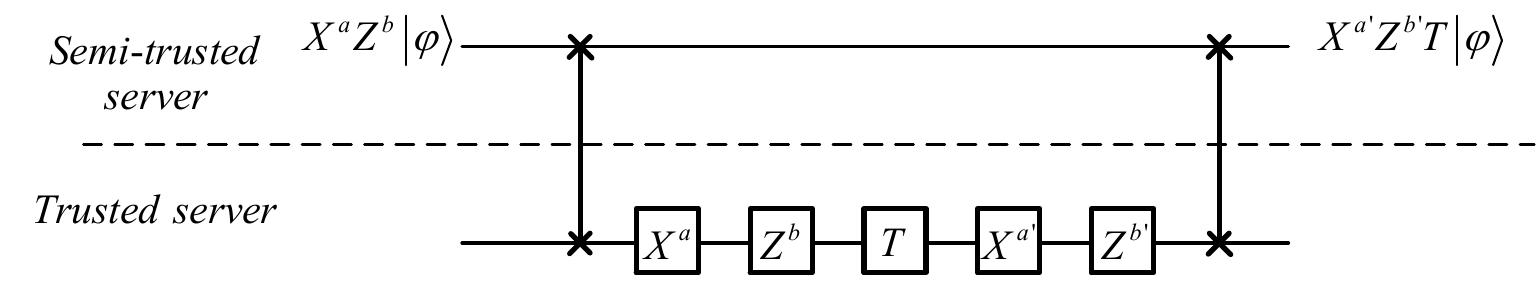}
		\caption{Update of $T$-gate based on trusted server after improvement.}
		\label{figure4}
	\end{figure}

	The client uploads the prepared quantum ciphertext data to the semi-trusted server and sends the key to the trusted server at the same time. The semi-trusted server builds the quantum circuit according to the requirements of the client. When there is a $T$-gate in the computing operation circuit, the quantum state of the $T$-gate to be executed is sent to the trusted server, and the trusted server decrypts the quantum state, then execute the $T$-gate on the plaintext state. Then a key is randomly generated to encrypt the quantum state, and the ciphertext is sent to the semi-trusted server, and the semi-trusted server continues to perform the calculation process. The specific processed quantum circuit is shown in the Fig.\ref{figure4}.
	
	The trusted server performs the corresponding key update process. When the calculation is completed, the semi-trusted server performs the measurement operation and returns the result to the client. After the trusted server is updated, it sends a new key to the client safely, and the client uses the key to decrypt the measurement result to obtain the real result.
	
	\section{Quantum k-means algorithm based on quantum cloud computing}
	\label{4}
\begin{figure}
		\includegraphics[width=0.9\linewidth]{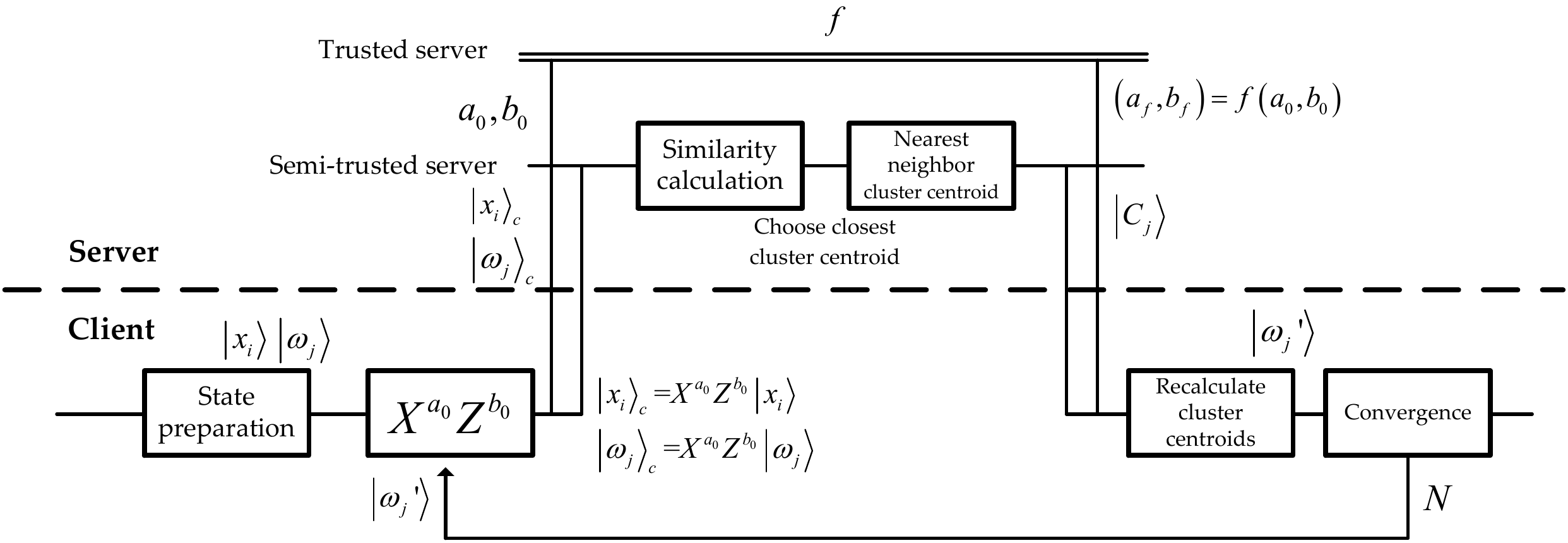}
		\caption{The process of quantum k-means algorithm based on quantum cloud computing.}
		\label{figure5}
	\end{figure}

	The classic k-means algorithm has been widely used in various fields since its appearance. In the era of big data, in order to solve the computational efficiency problem of the data clustering process when the data volume of the k-means algorithm is large, the quantum k-means algorithm changes some steps in the classical algorithm into a quantum version to accelerate. i.e. Use SwapTest to calculate the fidelity between quantum information, and replace the similarity between data points in classical k-means with the fidelity of quantum information; In addition, use GroverOptim to find the minimum value of the saved similarity information as the cluster center. For clients with weak quantum computing power, the iterative calculation of SwapTest and GroverOptim still has a serious burden. Therefore, we designed the quantum k-means algorithm based on quantum cloud computing as shown in Fig.\ref{figure5}.
	
	The data is encrypted at the client, the ciphertext is sent to the semi-trusted server, and the key is sent to the trusted server. The semi-trusted server executes $SwapTest$ and $GroverOptim$ operations, and the trusted server performs corresponding key updates according to the calculation process. When the calculation is completed, the semi-trusted server sends the calculated ciphertext result to the client. The trusted server sends the updated key to the client, and the client decrypts the calculation result and judges whether it has converged. The detailed steps of the algorithm are as follows:
	
	Suppose that a set of $M$ training data ${x_i},i \in \left\{ {1,...,M} \right\}$ is divided into $k$ sets, and the cluster centroid ${\omega _1},{\omega _2},...,{\omega _k}$ are randomly selected.
	\begin{framed}
		\centerline {\textbf{Quantum k-means algorithm based on quantum cloud computing}}
		\vbox{}
		\noindent\textbf{1.Initialize}\\ Training data ${x_i},i \in \left\{ {1,...,M} \right\}$ are prepared into quantum state $\left| {{x_i}} \right\rangle $, And initialize cluster centroids ${\omega _j},j \in \left\{ {1,...,k} \right\}$ to prepare a quantum state $\left| {{\omega _j}} \right\rangle $.
		
		\noindent\textbf{2.Encryption}\\ The client uses the $H$ gate to prepare the training data $\left| {{x_i}} \right\rangle $ and the cluster centroid $\left| {{\omega _j}} \right\rangle $ into a superposition state. Then randomly generate a 2n-bit key $ek = {\left( {{a_0},{b_0}} \right)^n}$ from $\left\{ {0,1} \right\}$, and use the key $ek$ to encrypt the plaintext superposition state to obtain the ciphertext superposition state ${\left| {{x_i}} \right\rangle _c}$,${\left| {{\omega _j}} \right\rangle _c}$. And send the ciphertext superposition state to the semi-trusted server, and send the key to the trusted server.
		
		\noindent\textbf{3.Choose closest cluster centroid} (Calculate i times in the server loop)
		\begin{itemize}
			\item[$ \bullet $] Calculate the value of $\left\| {{x_i} - {\omega _j}} \right\|$: The similarity between the $i$ th quantum state ${\left| {{x_i}} \right\rangle _c}$ and the $k$ cluster centroids ${\left| {{\omega _j}} \right\rangle _c}$ is calculated by SwapTest and stored in the quantum state $\left| \varphi  \right\rangle $.
			\item[$ \bullet $] Use GroverOptim to mark the similarity quantum state $\left| \varphi  \right\rangle $ and find the nearest cluster centroid ${\left| {{\omega _j}} \right\rangle _c}$ of the training data ${\left| {{x_i}} \right\rangle _c}$, And put ${\left| {{x_i}} \right\rangle _c}$ into the set ${C_j}$.
			\begin{equation}
			{C_j} = \left\{ {\mathop {\arg \min }\limits_k {{\left\| {{x_i} - {\omega _k}} \right\|}^2}} \right\}
			\end{equation}	
			The semi-trusted server returns the ciphertext result to the client, and the trusted server sends the updated decryption key $dk = \left( {{a_f},{b_f}} \right) \in {\left\{ {0,1} \right\}^n}$ to the client.	
		\end{itemize}
		\noindent\textbf{4.Calculate new cluster centroids} (The client needs to repeat the calculation $j$ times)
		After decryption at the client, recalculate the new cluster center ${\omega _j}'$ for the already allocated set ${C_j}$.
		\begin{equation}
		{\omega _j}' = \frac{1}{n}\sum\limits_{i \in {C_j}} {{x_i}}
		\end{equation}
		Where $n$ is the number of elements in set ${C_j}$. e.g. Suppose there are five elements $\left\{ {{x_2},{x_3},{x_6},{x_7},{x_8}} \right\}$ in set ${C_3}$. The new cluster centroid is calculated as ${\omega _3}' = \frac{1}{5}\left( {{x_2} + {x_3}{\rm{ + }}{x_6}{\rm{ + }}{x_7}{\rm{ + }}{x_8}} \right)$.
		
		\noindent\textbf{5.Convergence}\\
		The client sets the convergence threshold $\tau $ and calculates the difference between the centroid ${\omega _j}'$ of the $s$ iteration and the centroid ${\omega _j}$ of the $s-1$ iteration. When
		\begin{equation}
		\left\| {\omega _j^s - \omega _j^{s - 1}} \right\| < \tau
		\end{equation}
		is satisfied, the loop is terminated. Otherwise, the client repeats steps 3 and 4.
		
	\end{framed}
	As can be seen from the description, if the data scale $i$ is very large, it will put a lot of pressure on the client. In this paper, through the quantum homomorphic encryption scheme, the complex steps are uploaded to the server for calculation, and through the new T-gate update scheme to simplify the $T$-gate update process in the cloud. Next, we will verify the correctness of the calculation process in the quantum cloud.
	
	\section{Experiment of quantum subroutine SwapTest based on improved QHE scheme }
	\label{5}
	As a practical subroutine, the quantum subroutine SwapTest is widely used in all kinds of quantum machine learning to calculate the similarity between two quantum states instead of the linear distance between two quantum states. In this chapter, we will give the experiments of SwapTest in plaintext and ciphertext to verify the correctness.
	
\begin{figure}
		\includegraphics[width=0.5\linewidth]{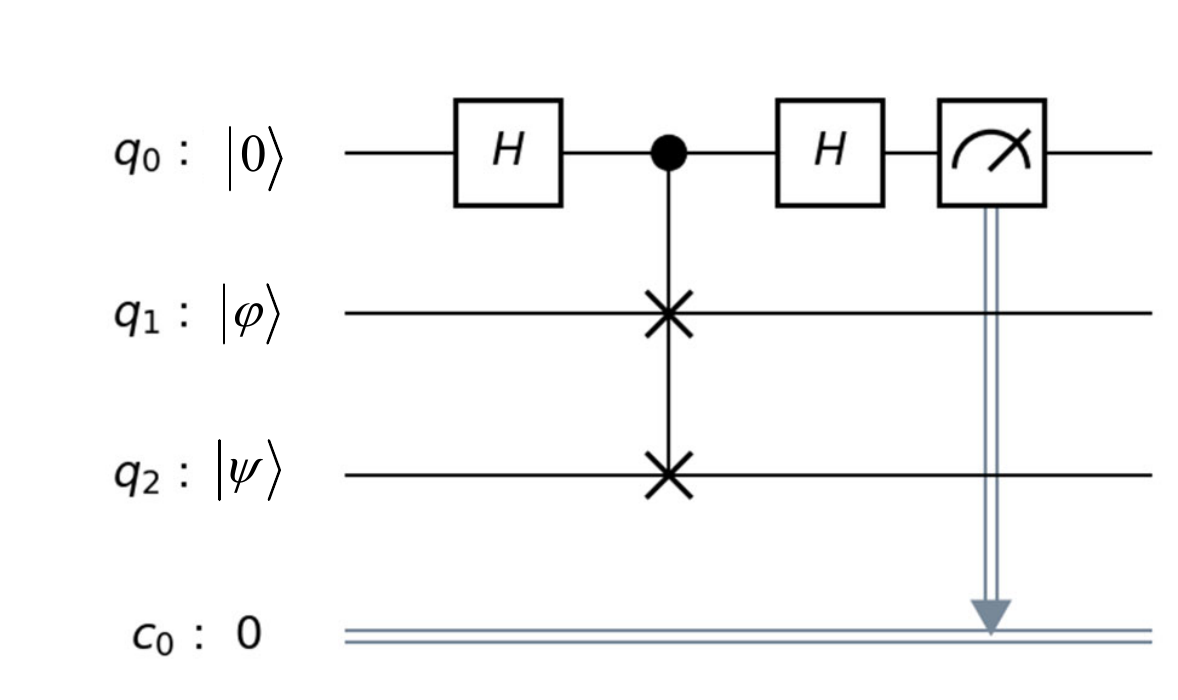}
		\caption{Quantum circuit of quantum subroutine SwapTest.}
		\label{figure6}
	\end{figure}

	In Step3, calculate the similarity between the two quantum states of ${\left| {{x_i}} \right\rangle _c}$ and ${\left| {{\omega _j}} \right\rangle _c}$ through the quantum subroutine SwapTest to indicate the distance $\left\| {{x_i} - {\omega _j}} \right\|$ between the data point and the cluster centroid. This method in Ref\cite{aimeur2006machine(4)} represents the similarity between quantum states by measuring the probability $P\left( {\left| 0 \right\rangle } \right)$ that the control bit is $\left| 0 \right\rangle$, instead of the distance between the classical vectors. The quantum circuit can be as shown in Fig.\ref{figure6}.
	
	In the quantum circuit, ${q_0}$ is the control bit, ${q_1},{q_2}$ is the input qubit, and the system state is $\left| {{\upsilon _1}} \right\rangle {\rm{ = }}\left| {0,\varphi ,\psi } \right\rangle $ at the time of input, and the state of the circuit becomes 
	\begin{equation}
	\left| {{\upsilon _2}} \right\rangle  = \frac{1}{2}\left| 0 \right\rangle \left( {\left| {\varphi ,\psi } \right\rangle  + \left| {\psi ,\varphi } \right\rangle } \right) + \frac{1}{2}\left| 1 \right\rangle \left( {\left| {\varphi ,\psi } \right\rangle  - \left| {\psi ,\varphi } \right\rangle } \right)
	\label{eq5}
	\end{equation}
	 after the circuit is executed. At this time, the probability $P\left( {\left| 0 \right\rangle } \right)$ when ${q_0}$ is $\left| 0 \right\rangle $ is measured.
	\begin{equation}
	\begin{array}{l}\vspace{1ex}
	P\left( {\left| 0 \right\rangle } \right) = {\left| {\frac{1}{2}\left\langle {0|0} \right\rangle \left( {\left| {\varphi ,\psi } \right\rangle  + \left| {\psi ,\varphi } \right\rangle } \right) + \frac{1}{2}\left\langle {0|1} \right\rangle \left( {\left| {\varphi ,\psi } \right\rangle  - \left| {\psi ,\varphi } \right\rangle } \right)} \right|^2}\\\vspace{1ex}
	\quad \quad \quad  = \frac{1}{4}{\left| {\left( {\left| {\varphi ,\psi } \right\rangle  + \left| {\psi ,\varphi } \right\rangle } \right)} \right|^2}\\
	\quad \quad \quad  = \frac{1}{2} + \frac{1}{2}{\left| {\left\langle {\varphi |\psi } \right\rangle } \right|^2}
	\end{array} 
	\end{equation}
	
\begin{figure}
		\includegraphics[width=0.7\linewidth]{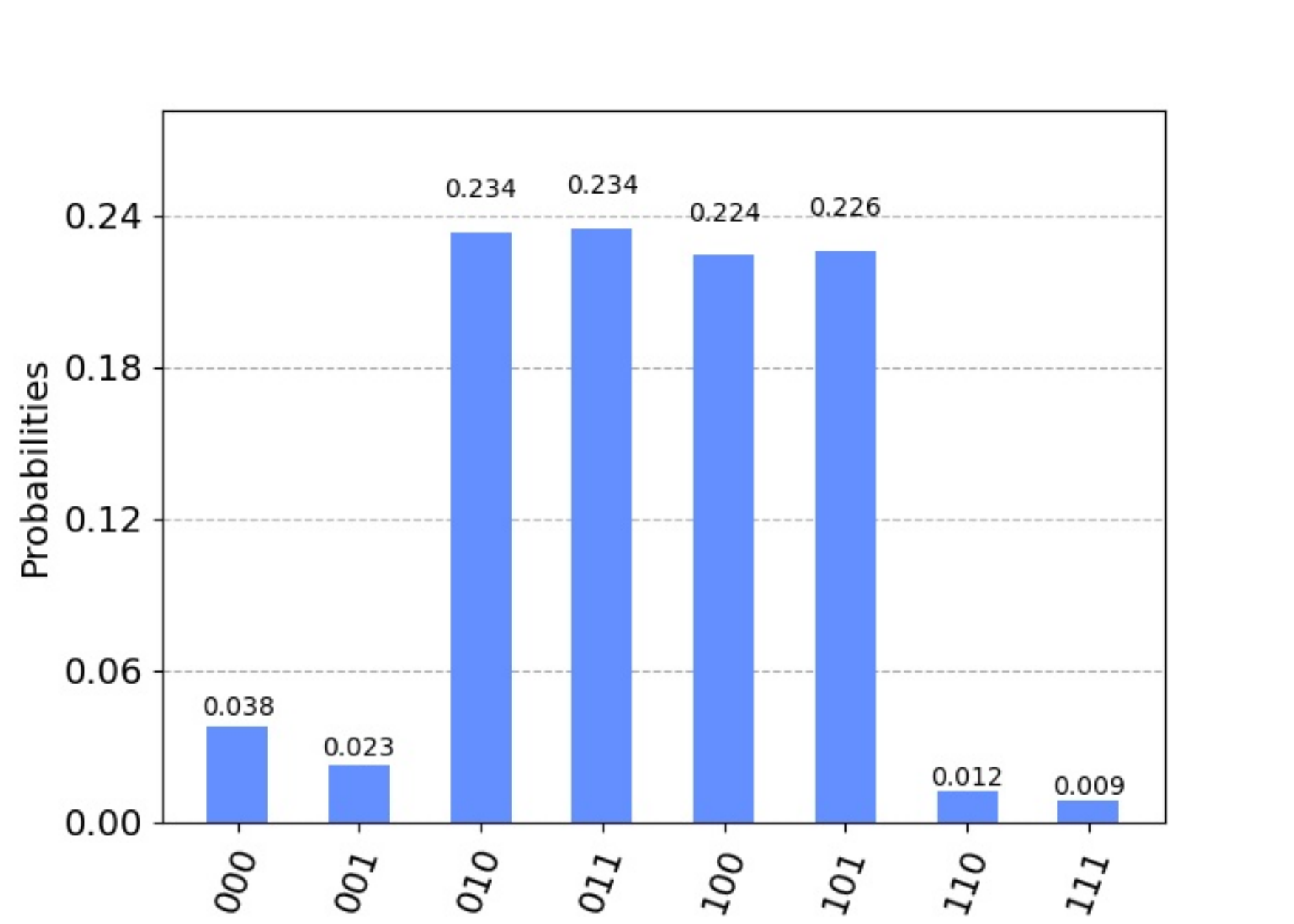}
		\caption{The result of measuring the similarity between $\left| 1 \right\rangle$ ,$\left| 0 \right\rangle $ in plaintext.}
		\label{figure7}
	\end{figure}

	It can be seen from Eq.\ref{eq5} that the greater the similarity between $\left| \varphi  \right\rangle $ and $\left| \psi  \right\rangle $, the closer the probability $P\left( {\left| 0 \right\rangle } \right)$ is to 1. On the contrary, the probability $P\left( {\left| 0 \right\rangle } \right)$ is closer to 0.5. Suppose we want to measure the similarity between $\left| 0 \right\rangle $ and $\left| 1 \right\rangle $. Obviously $\left| 0 \right\rangle $ and $\left| 1 \right\rangle $ are orthogonal, and the probability is $P\left( {\left| 0 \right\rangle } \right) = 0.5$. The circuit in Fig.\ref{figure6} is executed 8192 times on IBM Qiskit, and the measurement results obtained in the environment containing noise are shown in Fig.\ref{figure7}.
	
\begin{figure}
		\includegraphics[width=0.9\linewidth]{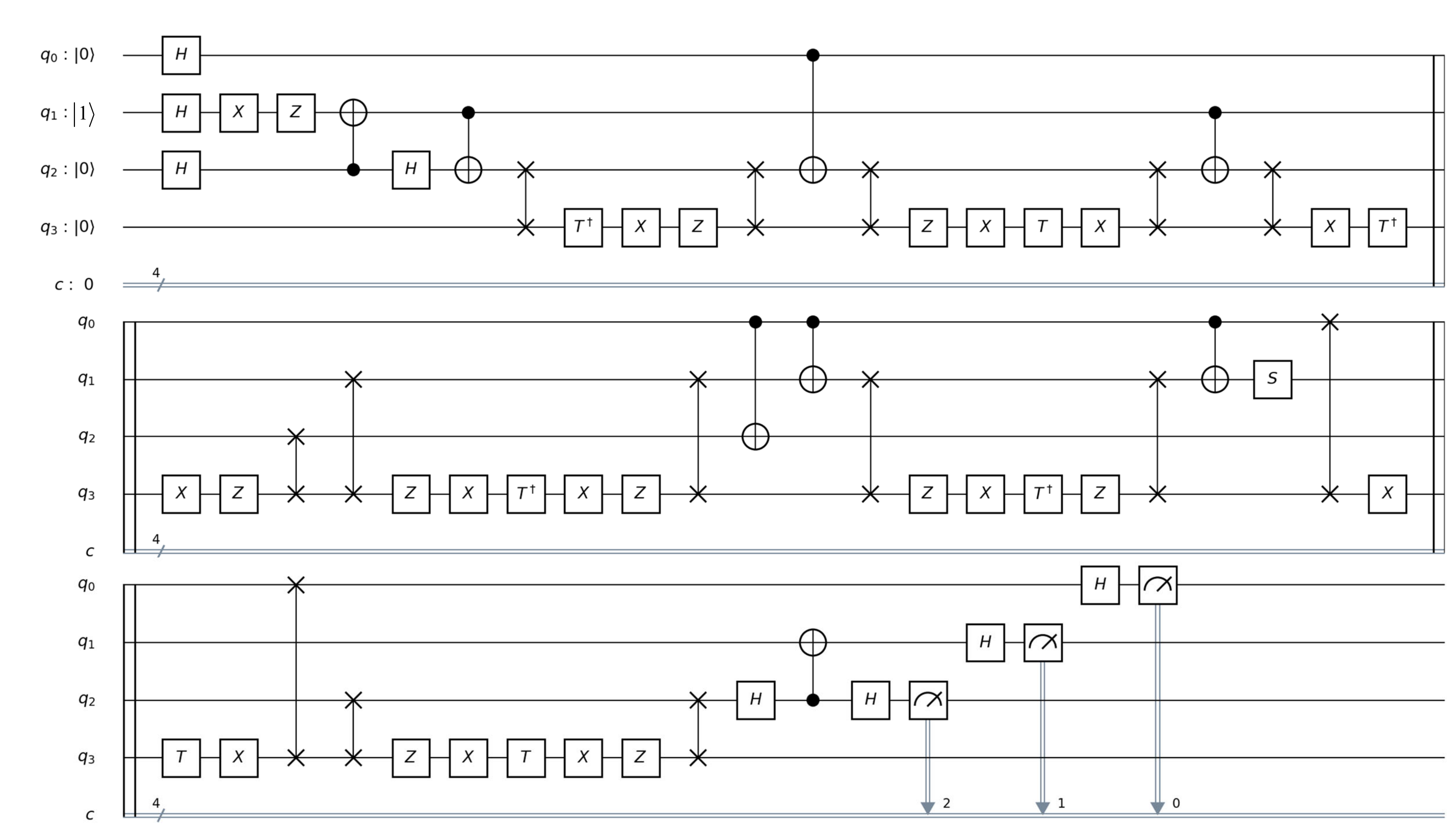}
		\caption{The server ciphertext executes the quantum circuit of measuring $\left| 1 \right\rangle ,\left| 0 \right\rangle $ similarity by SwapTest.}
		\label{figure8}
	\end{figure}

	Because of the noise and imperfect quantum gates in the circuit, the results we measured $P\left( {\left| 0 \right\rangle } \right) = 0.038{\rm{ + }}0.234 + 0.224{\rm{ + }}0.012 = 0.508 \approx 0.5$. So it can be proved that they are orthogonal. In the case of encryption, we design the specific quantum circuit of the quantum subroutine SwapTest in the server. Consider the situation under ciphertext according to the improved homomorphic encryption scheme in Sect.\ref{3}, and replace the Toffoli gate with the combination of $H$ gate, $S$ gate, $T$ gate and $CNOT$ gate, and use the key update process of each door to update the key. The specific circuit is shown in Fig.\ref{figure8}. 
	
	In the quantum circuit, ${q_0},{q_1},{q_2}$ is the quantum register that executes SwapTest in ciphertext on the semi-trusted server. The encryption key of ${q_1},{q_2}$ is $\left\{ {1,1} \right\},\left\{ {0,0} \right\}$ respectively, and ${q_0}$ is the auxiliary qubit to be measured to store the value of similarity, so there is no need for encryption. ${q_3}$ is the trusted server used to assist the semi-trusted server to perform T-gate key update operation and update the key. The circuit is also executed on IBM Qiskit for 8192 times, and the measurement results are shown in the following Fig.\ref{figure9}.
	
	In this result, we can see the probability that ${q_0}$ is 0 under the ciphertext. $P\left( {\left| 0 \right\rangle } \right){\rm{ = }}0.055{\rm{ + }}\\0.209{\rm{ + }}0.201{\rm{ + }}0.042{\rm{ = }}0.507 \approx 0.5$. Comparing the results of the similarity between SwapTest in the ciphertext and it in the plaintext, we can clearly see that when we encrypt the input qubits in, the similarity results obtained are approximately equal. 
	\begin{figure}
		\includegraphics[width=0.6\linewidth]{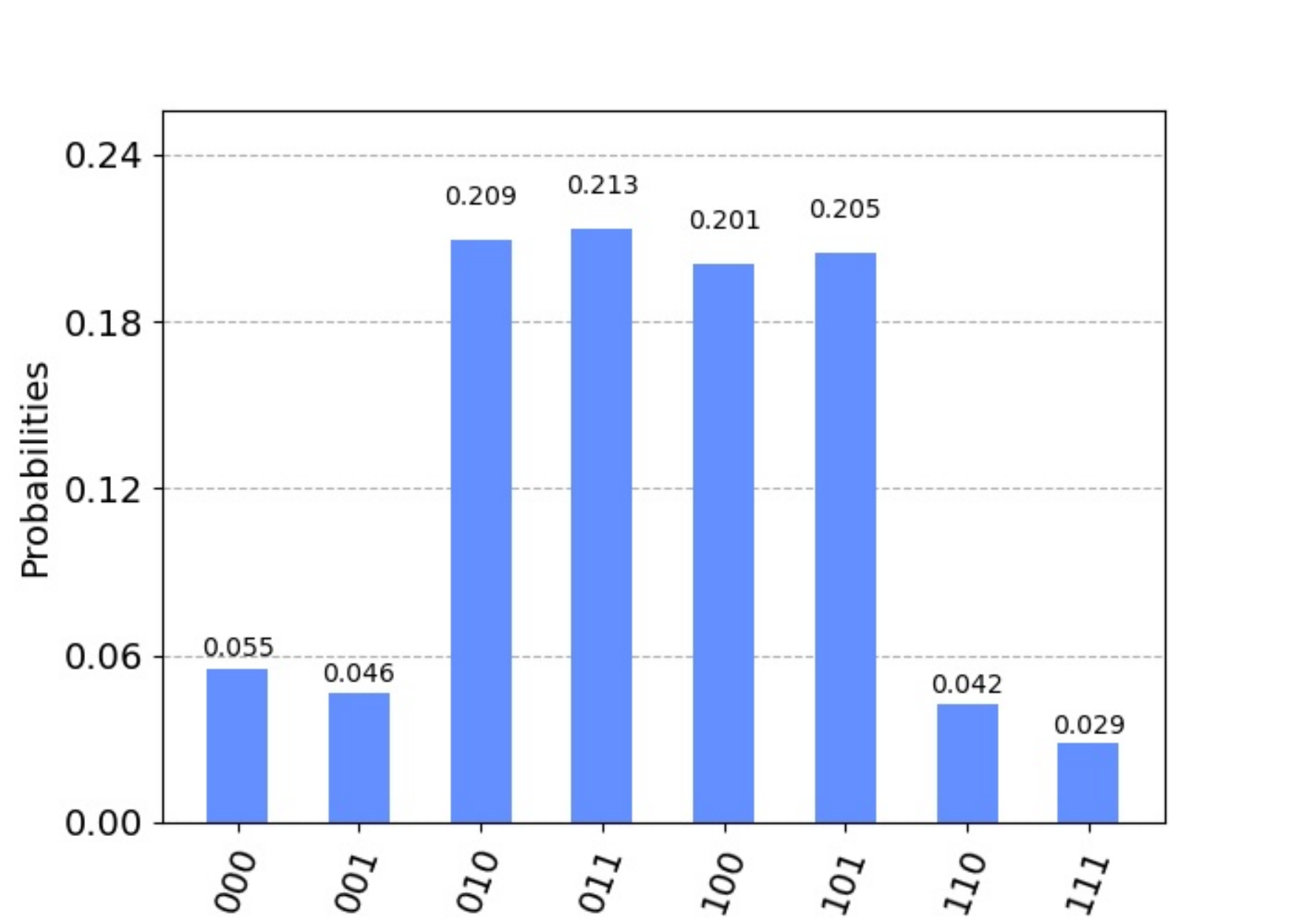}
		\caption{The server ciphertext executes the quantum circuit of measuring $\left| 1 \right\rangle ,\left| 0 \right\rangle $ similarity by SwapTest.}
		\label{figure9}
	\end{figure}
	
	Therefore our scheme is correct and effective. In addition, we didn't update the key here to get the result, because the result we finally need is not encrypted in the register. But for the client and server, as long as the quantum state measured in ${q_1},{q_2}$ is encrypted, security can be guaranteed.
	
	Finally, we store the calculated $k$ values $\left\| {{x_i} - {\omega _j}} \right\|$ in the quantum state $\left| \lambda  \right\rangle $ through phase estimation for the next calculation.
	
	\section{Experiment of Quantum minimization algorithm based on improved QHE scheme}
	\label{6}
	Quantum minimization algorithm\cite{durr1996quantum(9)} as an extension of the Grover algorithm is widely used in various quantum machine learning algorithms, also known as $GroverOptim$, and is one of the core steps of the quantum k-means algorithm.
	
	In this algorithm, the quantum state $\left| \lambda  \right\rangle $ calculated by SwapTest is regarded as a search of disordered database by using Grover search algorithm. Through the parallelism of quantum computing, the optimal solution of the problem can always be found after several iterations, which is very important for classical machine learning which is always easy to fall into local optimun. However, these operations need to be repeated in the actual computing process, which is very cumbersome for the client, so we can achieve it in the quantum cloud computing. This section will focus on the implementation of the algorithm in the ciphertext state of the improved quantum homomorphic encryption (QHE) scheme.
	
\begin{figure}
		\includegraphics[width=0.6\linewidth]{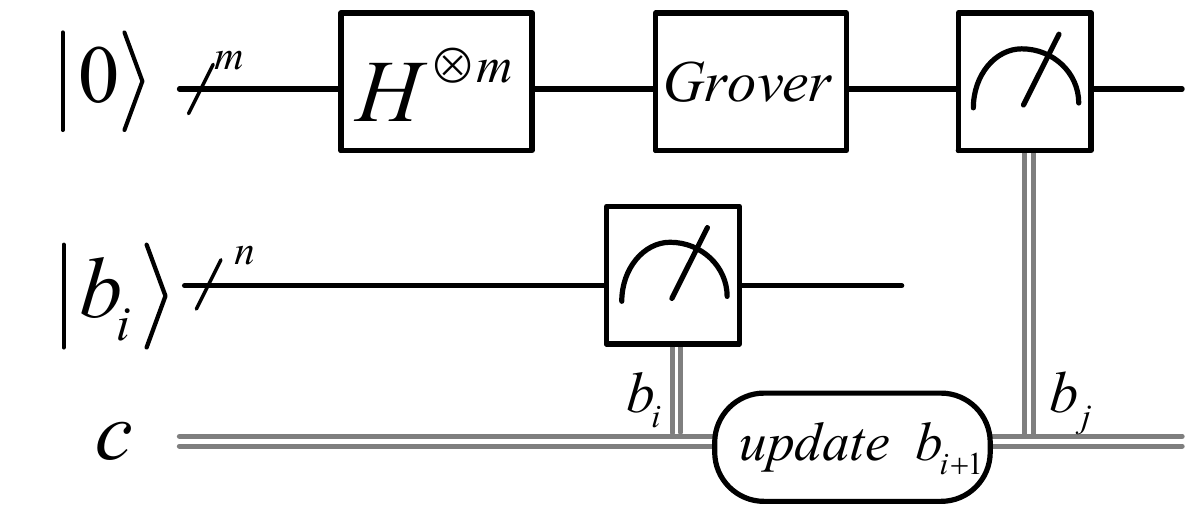}
		\caption{Quantum Circuit of Quantum minimization algorithm.}
		\label{figure10}
	\end{figure}

	We will briefly describe the specific steps of GroverOptim, which requires $m$ registers to store the input qubit $\left| 0 \right\rangle $ (where ${2^m} = k$), $n$ qubit storage threshold $\left| {{b_i}} \right\rangle $. The quantum circuit of the algorithm is shown in Fig.\ref{figure10}.
	
	\begin{framed}
		\centerline {\textbf{Quantum minimization algorithm based on quantum cloud computing}}
		\vbox{}
		\noindent\textbf{1.Initialize}\\Define the $a$-th value in the similarity quantum state $\left| \lambda  \right\rangle $ as $f\left( a \right)$ (in this case, both $a$ and $f\left( a \right)$ are expressed in binary form). And randomly select ${a_1}$ and ${b_1} = f\left( {{a_1}} \right)$ as the threshold for the first iteration.
		
		\noindent\textbf{2.Apply Hadamard gate}\\Apply the Hadamard gate to the first register, so that the quantum states in the register are superimposed to represent all possible states in ${a_i}$.
		
		\noindent\textbf{3.Apply Grover search algorithm}\\Apply Quantum Oracle to mark all states with a value less than ${b_i}$. Choose one as the output $\left| {{a_j}} \right\rangle $ of the algorithm and convert it to the input ${a_j}$ to calculate the value of ${b_j}{\rm{ = }}f\left( {{a_j}} \right)$.
		
		\noindent\textbf{4.Update threshold}\\Compare the result ${b_j}$ with the threshold ${b_i}$. If ${b_j} > {b_i}$, then set ${a_{i + 1}} = {a_j}$ and ${b_{i + 1}} = {b_j}$. Otherwise, set ${a_{i + 1}} = {a_i}$,${b_{i + 1}} = {b_i}$. 
		
		When the algorithm iterates to $\sqrt {{2^m}} $ times, the algorithm is terminated. Measure the state $\left| {{a_{\min }}} \right\rangle $ in the register and calculate the minimum value ${b_{\min }}$.
	\end{framed}
	
	Obviously, the parallelism of the algorithm is mainly attributed to the Grover search algorithm, which applies the gate operation to the superposition state of all possible inputs. Because of this, the quantum minimization algorithm can always find the global optimal solution. The number of iterations x of the algorithm is calculated accurately in\cite{durr1996quantum(9),ahuja1999quantum(25)}. As $m$ increases, the speed of the algorithm increases. Similarly, when the client's quantum computing power is limited, the larger $m$ means the larger the number of registers required. Therefore, it is very necessary to use $GroverOptim$ in quantum cloud computing. Next, an example is used to verify the correctness of $GroverOptim$ algorithm on ciphertext.
	
\begin{figure}
		\includegraphics[width=0.6\linewidth]{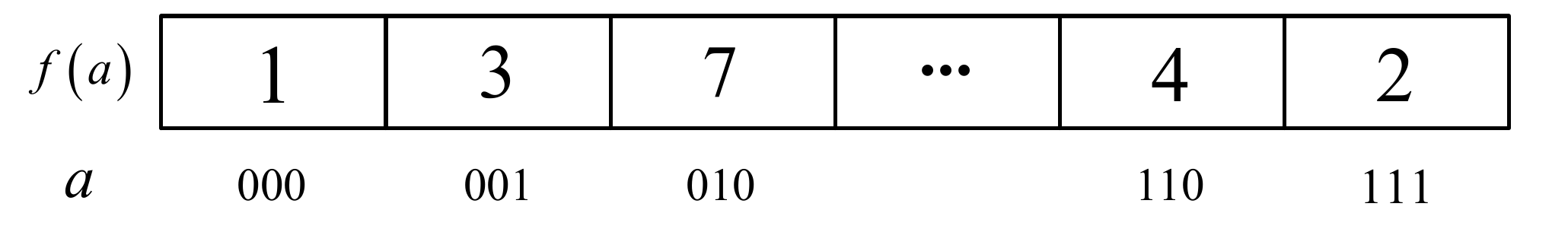}
		\caption{Correspondence of unordered databases $f\left( a \right)$.}
		\label{figure11}
	\end{figure}

	Suppose there is an unordered database $f\left( a \right)$, which contains 8 data, and the corresponding relationship is shown in Fig.\ref{figure11}.
	
	Then part of the data in the database can be expressed as 
	\[f\left( a \right) = \left\{ \begin{array}{l}
	1\quad {\kern 1pt} a = 000\\
	3\quad a = 001\\
	2\quad a = 111\\
	otherwise\quad 
	\end{array} \right.\]
	
	For instance, we randomly select the starting threshold value ${b_1} = 3,{a_1} = 001$, so we design a quantum Oracle to mark those less than the threshold ${b_1}$.
	
	\begin{equation}
	O\left| a \right\rangle  = {\left( { - 1} \right)^{h\left( a \right)}}\left| a \right\rangle ,\quad h\left( x \right) = \left\{ \begin{array}{l}
	0\quad f\left( a \right) > 1\\
	1\quad f\left( a \right) < 1
	\end{array} \right.
	\end{equation}

\begin{figure}
		\includegraphics[width=0.5\linewidth]{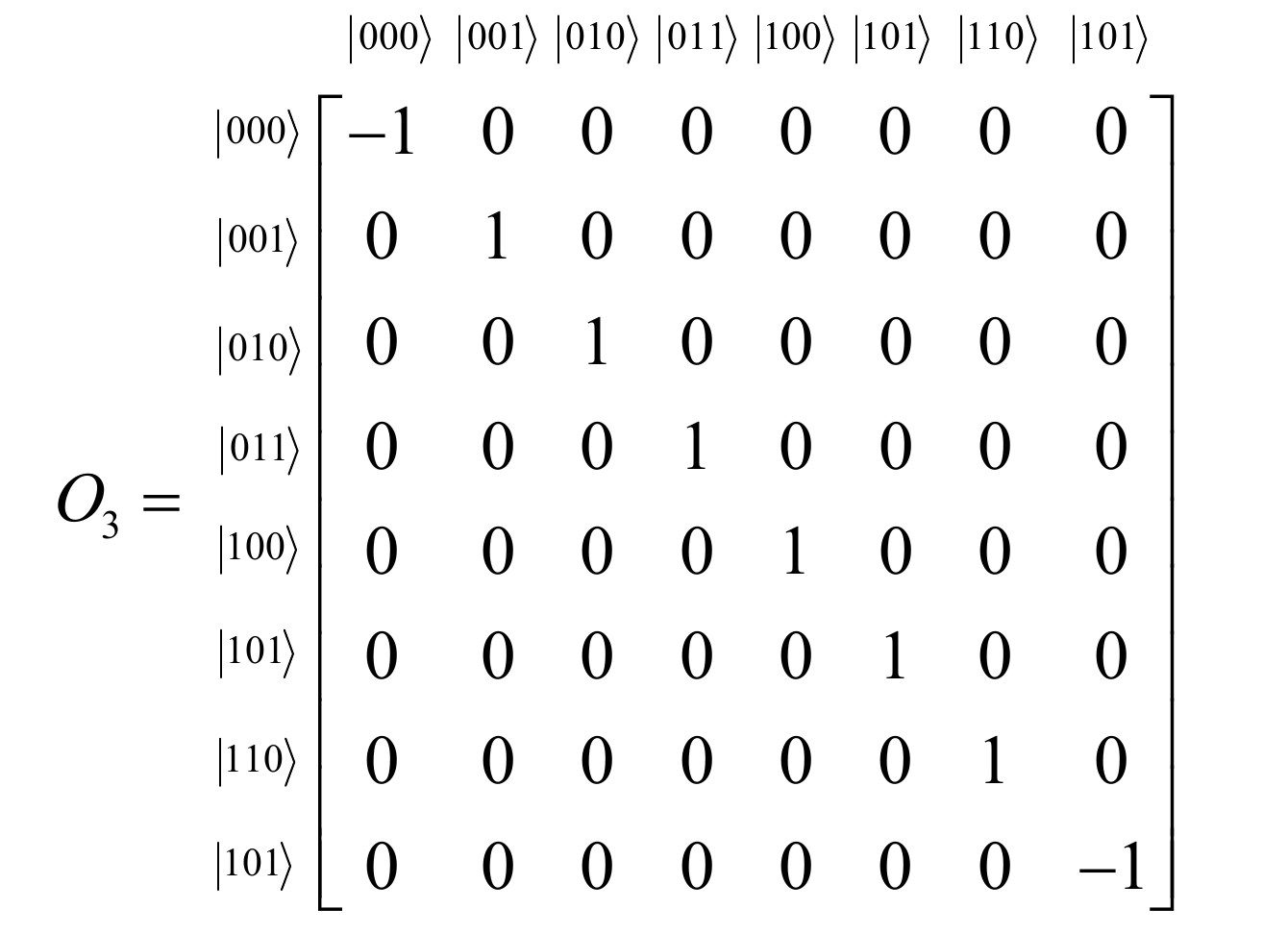}
		\caption{The Matrix form of Quantum Oracle ${O_3}$.}
		\label{figure12}
	\end{figure}

	At this point, the quantum Oracle ${O_3}$ marks the value $a = \left| {000} \right\rangle ,a = \left| {111} \right\rangle $ of all ${a_i}$ of $f\left( a \right) < {b_1}$, and the matrix form of the ${O_3}$ is shown in the Fig.\ref{figure12}.
	
\begin{figure}
		\includegraphics[width=0.9\linewidth]{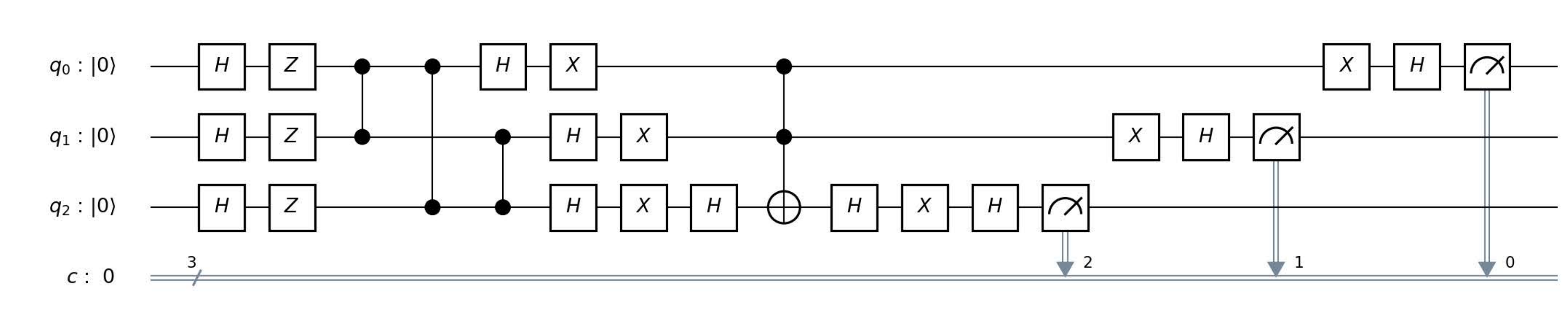}
		\caption{Searching for quantum circuits of $\left| {000} \right\rangle ,\left| {111} \right\rangle $ on Plaintext.}
		\label{figure13}
	\end{figure}

	The circuit that uses $Grover$ search algorithm to search for $\left| {000} \right\rangle ,\left| {111} \right\rangle $ in plaintext can be realized by the quantum circuit in Fig.\ref{figure13}.
	
\begin{figure}
		\includegraphics[width=0.6\linewidth]{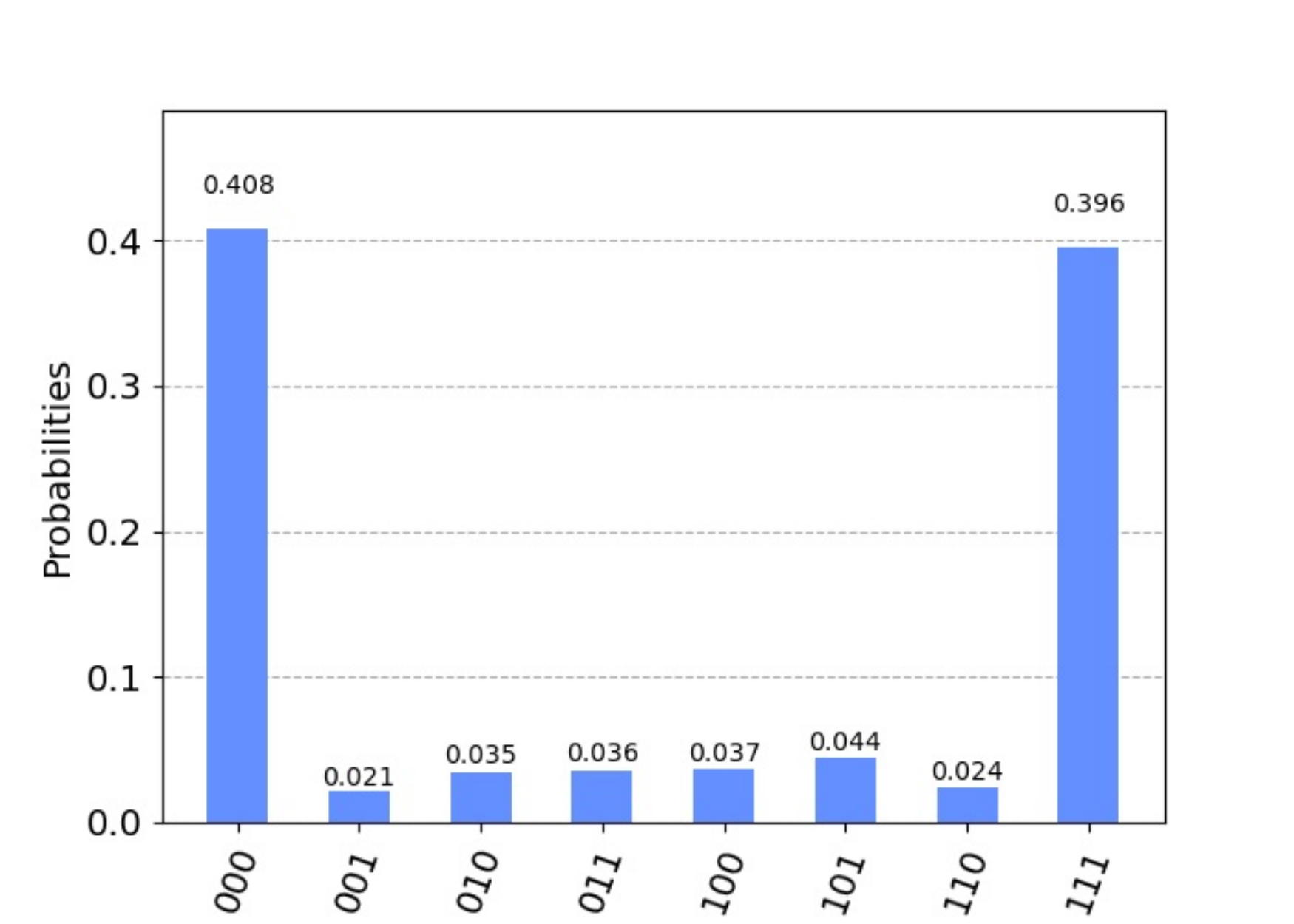}
		\caption{Searching for quantum circuits of $\left| {000} \right\rangle ,\left| {111} \right\rangle $ on Plaintext.}
		\label{figure14}
	\end{figure}

	The measurement results of the quantum circuit performed 8192 times in the IBM Qiskit simulation environment are shown in Fig.\ref{figure14}.
	
	
\begin{figure}
		\includegraphics[width=0.9\linewidth]{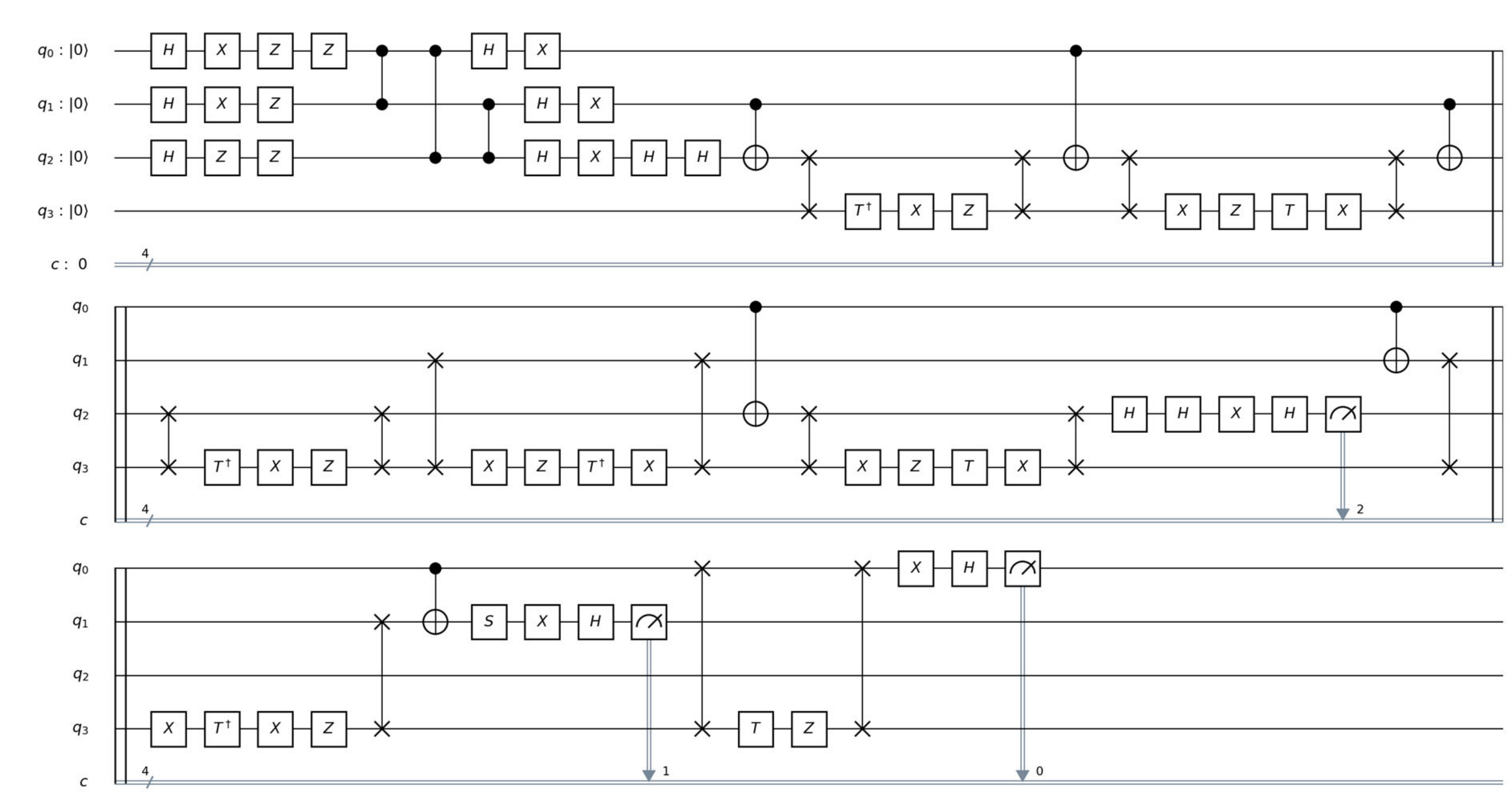}
		\caption{Searching for quantum circuits of $\left| {000} \right\rangle ,\left| {111} \right\rangle $ on Plaintext.}
		\label{figure15}
	\end{figure}

	According to the experimental results, it can be seen that the state $\left| {000} \right\rangle ,\left| {111} \right\rangle $ is obtained with a very high probability, indicating that the circuit can correctly search the corresponding quantum state in plaintext. In the case of encryption, assuming that the client’s initial encryption key is $\left\{ {1,1} \right\},\left\{ {0,1} \right\}$ and $\left\{ {0,1} \right\}$,The key means the first qubit uses ${X^1}{Z^1}$ to encrypt, and the second qubit uses ${X^0}{Z^1}$ to encrypt, the third qubit uses ${X^0}{Z^1}$ to encrypt. The Toffoli gate also uses the combination of $H$ gate, $S$ gate, $T$ gate and $CNOT$ gate to replace the form. we give the specific circuit and operation process. The circuit is shown in Fig.\ref{figure15}.
	
	The difference between the execution of the SwapTest circuit under the ciphertext to find the similarity is that the client needs to decrypt the results after the search is executed, and the key update process is completed by the trusted server and return the final key to the client. In order to verify that we found the desired result. According to the previous key update rule, we give the key update process of Grover searching for $\left| {000} \right\rangle ,\left| {111} \right\rangle $. As shown in the table\ref{table1}.
	\begin{table}
		\caption{The server searches for the key update process on the ciphertext}
		\label{table1}
		\scalebox{1}{
			\begin{tabular}{ccccccccc}
				\toprule
				\midrule
				& Initial key & $C{Z_{0,1}}$ & $C{Z_{0,2}}$ & $C{Z_{1,2}}$ & ${H_{0,1,2}}$ & ${H_2}$ & ${H_2}$ & $CNO{T_{1,2}}$\\
				\midrule
				${q_0}$ & $\left\{ {1,1} \right\}$ & $\left\{ {1,0} \right\}$ & $\left\{ {1,0} \right\}$ & & $\left\{ {0,1} \right\}$ & & & \\
				${q_1}$ & $\left\{ {1,0} \right\}$ & $\left\{ {1,1} \right\}$ & & $\left\{ {1,1} \right\}$ & $\left\{ {1,1} \right\}$ & & &$\left\{ {1,1} \right\}$\\
				${q_2}$ & $\left\{ {0,1} \right\}$ & & $\left\{ {0,0} \right\}$ & $\left\{ {0,1} \right\}$ & $\left\{ {1,0} \right\}$ & $\left\{ {0,1} \right\}$ & $\left\{ {1,0} \right\}$ & $\left\{ {0,0} \right\}$\\
				\bottomrule
				& $T_2^\dag $ & $CNO{T_{0,2}}$ & ${T_2}$ & $CNO{T_{1,2}}$ & $T_2^\dag $ & $CNO{T_{0,2}}$ & $T_1^\dag $ & $CNO{T_{0,1}}$  \\
				\midrule
				${q_0}$ & & $\left\{ {0,0} \right\}$ & & & & $\left\{ {0,1} \right\}$ & $\left\{ {0,1} \right\}$ \\
				${q_1}$ & & & & $\left\{ {1,1} \right\}$ & & & $\left\{ {1,0} \right\}$ & $\left\{ {1,0} \right\}$                    \\
				${q_2}$ & $\left\{ {1,1} \right\}$ & $\left\{ {1,1} \right\}$ & $\left\{ {1,0} \right\}$ & $\left\{ {0,0} \right\}$ & $\left\{ {1,1} \right\}$ & $\left\{ {1,1} \right\}$ & &                    \\
				\toprule
				& ${T_2}$ & $T_1^\dag $ & $CNO{T_{0,1}}$ & ${H_2}$ & ${T_1}$ & ${S_1}$ & ${H_2}$ & ${H_{0,1,2}}$ \\
				\midrule
				${q_0}$ & & & $\left\{ {0,0} \right\}$ & & $\left\{ {0,1} \right\}$ & & & $\left\{ {1,0} \right\}$ \\
				${q_1}$ & & $\left\{ {1,1} \right\}$ & $\left\{ {1,1} \right\}$ & & & $\left\{ {1,0} \right\}$ & & $\left\{ {0,1} \right\}$\\
				${q_2}$ & $\left\{ {1,0} \right\}$ & & & $\left\{ {0,1} \right\}$ & & & $\left\{ {1,0} \right\}$ & $\left\{ {0,1} \right\}$ \\
				\midrule
				\bottomrule
			\end{tabular}
		}
	\end{table}
	
\begin{figure}
		\includegraphics[width=0.6\linewidth]{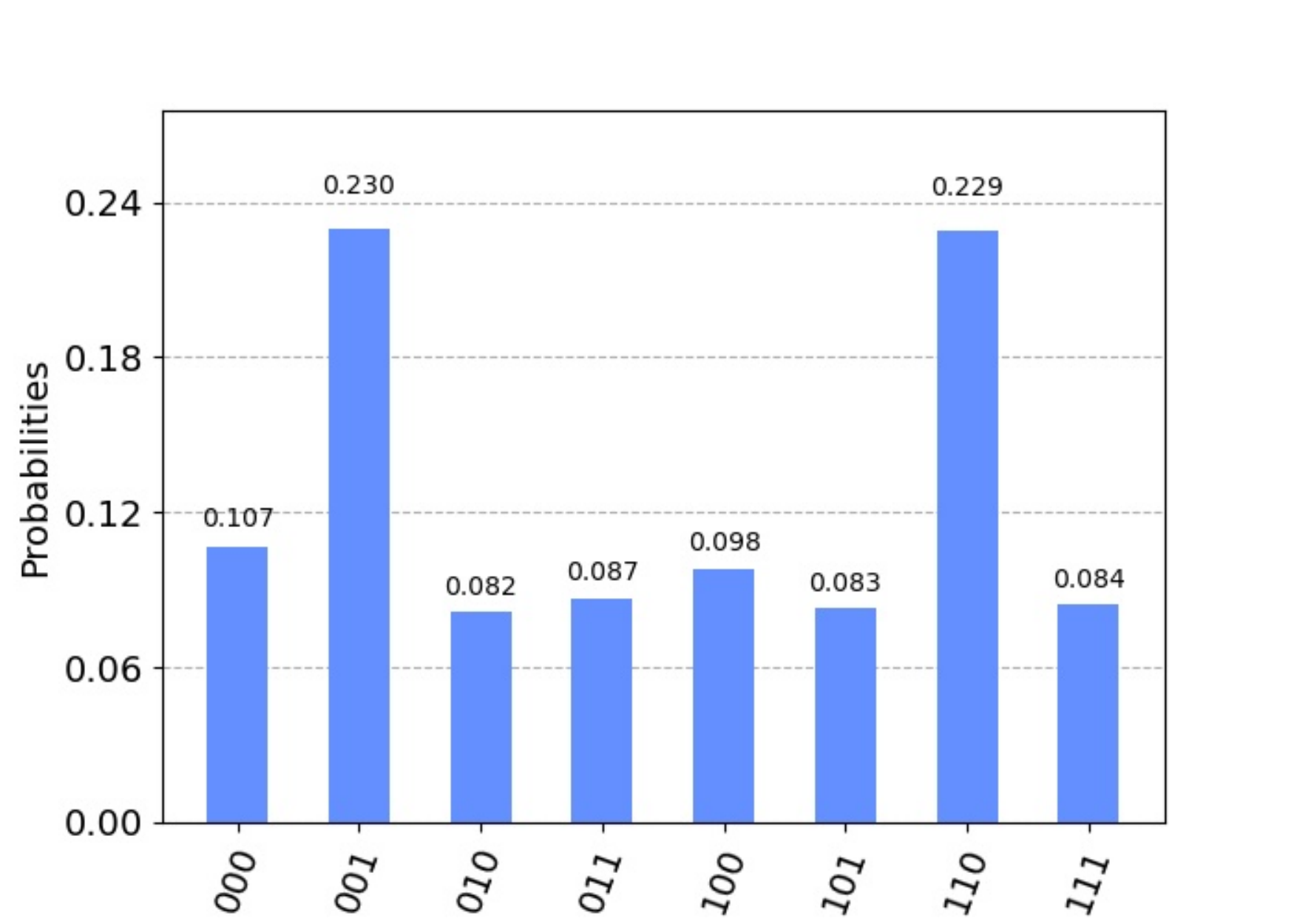}
		\caption{The server executes the ciphertext search for $\left| {000} \right\rangle ,\left| {111} \right\rangle $ experimental results.}
		\label{figure16}
	\end{figure}

	The final key ${\left\{ {{a_f},{b_f}} \right\}^3} = \left\{ {1,0} \right\},\left\{ {0,1} \right\},\left\{ {0,1} \right\}$ is calculated according to the key update rule. The key update process in the above table omits the update process of the X gate and the Z gate, but the update process includes them and only represents the key at that time when it reaches a certain quantum gate. For a single quantum bit gate, the subscript indicates the register to which the quantum gate is applied. e.g. ${H_{0,1,2}}$ indicates that the quantum gate apply on the ${q_0},{q_1},{q_2}$ quantum registers respectively. For the double qubit $CZ$ gate, the subscript represents the qubits of the gate. e.g. $C{Z_{0,1}}$ means that the $CZ$ gate is applied to the ${q_0},{q_1}$ quantum registers, and the control bit and the target bit are not distinguished.  For the double qubit gates-$CNOT$ gates, we need to distinguish between the control qubit and the target qubit. $CNO{T_{1,2}}$ indicates that the CNOT gate acts on the first and second quantum registers, where the first qubit is the control qubit and the second qubit is the target qubit. The quantum circuit is executed 8192 times in the IBM Qiskit simulation environment, and the result is shown in Fig.\ref{figure16}.
	
	According to the results, it can be seen that the two quantum states of $\left| {100} \right\rangle $ and $\left| {011} \right\rangle $ are obtained with a higher probability after the circuit is executed. Use the key ${\left\{ {{a_f},{b_f}} \right\}^3}$ derived from the previous calculation to decrypt the two quantum states according to the decryption formula.
	\begin{equation}
	{\left| {{x_i}} \right\rangle _p} = {X^{{a_f}}}{Z^{{b_f}}}{\left| {{x_i}} \right\rangle _c}\quad {a_f},{b_f} \in \left\{ {0,1} \right\}
	\end{equation}
	Then the two quantum states are decrypted as
	\begin{equation}
	\left| {100} \right\rangle \xrightarrow{{\left\{ {1,0} \right\},\left\{ {0,1} \right\},\left\{ {0,1} \right\}}} \left| {000} \right\rangle,\quad \left| {011} \right\rangle \xrightarrow{{\left\{ {1,0} \right\},\left\{ {0,1} \right\},\left\{ {0,1} \right\}}} \left| {111} \right\rangle
	\end{equation}
	
	The result of decryption on ciphertext is the same as the result of plaintext search for $\left| {000} \right\rangle ,\left| {111} \right\rangle $ two quantum states, It can prove that the plan under our ciphertext is correct. 
	
	Because we searched for $\left| {000} \right\rangle ,\left| {111} \right\rangle $ two states that are less than the threshold ${b_i} = 3$. According to the algorithm we designed, we randomly select a state ${a_i} = \left| {111} \right\rangle $ and update the threshold ${b_{i + 1}} = f\left( {{a_i}} \right) = 2$ for the next iterative calculation.
	
	\section{Analysis}
	\label{7}
	This section will analyze our algorithm scheme in terms of the security of the quantum homomorphic encryption scheme based on the trusted server and the operational efficiency of the quantum algorithm. The influence of the noise in the experiment on the experiment will also be analyzed.
	
	When the classical k-means algorithm needs to iterate $t$ times when calculating $k$ clustering centers, $n$-dimensional data vectors and the scale of $M$. The time complexity of the algorithm is $O\left( {Mnkt} \right)$. While quantum k-means calculates the distance between the $n$-dimensional data vector and the clustering center $\omega $, the time complexity is reduced to $O\left( {\log \left( n \right)} \right)$ by subtly calculating the similarity of two high-dimensional vectors instead of the distance. In addition, in the quantum minimum search algorithm, when the number of clustering centers $k$ is very large. We represent all $k$ input ($\sqrt {{2^m}}  = k$) by using $m$ quantum states. Reduce the time complexity of searching for the minimum value to $O\left( {\sqrt k } \right)$. There is a secondary acceleration in the algorithm. In summary, the time complexity of the quantum algorithm can be expressed as $O\left( {M\log \left( n \right)\sqrt k t} \right)$. And when the value of $k$ and $n$ are larger, the importance of the client's use of quantum cloud computing due to its own weak quantum power becomes more prominent, and the acceleration effect of quantum algorithms becomes more prominent. In the case of client encryption and decryption, the operation performed by the client is related to the number of quantum states $m$ required. But for the overall algorithm, the operation performed by the client is polynomial.
	
	On the other hand, the ciphertext search scheme based on trusted server is secure. From the overall point of view, our scheme is compact, and the decryption process of the scheme is independent of the evaluation quantum circuit QC that performs the quantum computation. From the point of view of the scheme steps, first of all, in the encryption process, the client randomly generates the key and uses the quantum method to encrypt the plaintext, because our encryption and evaluation ($SwapTest$, $GroverOptim$) and other operations are applied to the totally mixed state, Therefore, this solution satisfies.
	\begin{equation}
	\frac{1}{{{2^{2n}}}}\sum\limits_{a,b \in {{\left\{ {0,1} \right\}}^n}} {{X^a}} {Z^b}\sigma {Z^b}{X^a} = \frac{{{I_{{2^n}}}}}{{{2^n}}}
	\end{equation}
	Where $\frac{{{I_{{2^n}}}}}{{{2^n}}}$ represents the totally mixed state, and $\sigma $ represents the density matrix form of the plaintext. It is completely consistent with the quantum one-time pad (QOTP)\cite{boykin2003optimal(16)} scheme, which has been verified to be correct. When executing to the T-gate, the semi-trusted server needs to send the quantum state to the trusted server, which is decrypted by the trusted server, and then directly performs the T-gate operation on the plaintext. Then the semi-trusted server generates the key randomly and encrypts it and uploads it to the semi-trusted server. The semi-trusted server will not get any information about the plaintext or the key, so the process is secure. It can be proved that our scheme satisfies the existence of quantum state ${\rho ^{E'}}$ so that all quantum states satisfy ${\psi ^{EU}}$.
	\begin{equation}
	\left\| {QHE.Enc\left( {{\psi ^{EU}}} \right) - {\rho ^{E'}} \otimes {\psi ^U}} \right\| = 0
	\end{equation}
	$QHE.Enc$ is the encryption algorithm executed on the encryption part of the input quantum state $\left| \psi  \right\rangle $, and ${\psi ^U}$ is the evaluation process of executing the search circuit. QHE is the set of quantum states formed by the probability distribution of the key and the randomness of the quantum state when performing the computing. It satisfies the definition of perfect security in QHE scheme\cite{liang2020teleportation(21)}. In summary, the ciphertext calculation scheme we researched guarantees security.
	
	It is worth noting that the experiment in this article is based on the IBM real quantum computing platform to simulate the noise environment in $ibmq\_16\_melbourne$. We will discuss the impact of noise on this scheme.
	In the first experiment, we measure the quantum register ${q_0}$ in the quantum circuit of SwapTest in plaintext and ciphertext respectively, and the probability of $\left| 0 \right\rangle $ is obtained.
	\[\begin{array}{l}
	{P_p}\left( {\left| 0 \right\rangle } \right){\rm{ = }}0.038{\rm{ + }}0.234 + 0.224{\rm{ + }}0.012 = 0.508\\
	{P_c}\left( {\left| 0 \right\rangle } \right){\rm{ = }}0.055{\rm{ + }}0.209{\rm{ + }}0.201{\rm{ + }}0.042{\rm{ = }}0.507
	\end{array}\]
	Obviously, both of them are approximately equal to 0.5, but there are some error in the experiment.  The main reason for this error (considering that there is no reading error) is the incompleteness of the $T$-gate. According to the matrix form of the basic quantum $T$-gate $T = \left[ {\begin{array}{*{20}{c}}
		1&0\\
		0&{{e^{i\pi /4}}}
		\end{array}} \right]$, it is obvious that every time we execute the $T$-gate, it is an approximate estimation operation. Therefore, when there is a certain scale of $T$-gates in the quantum circuit, the experiment has certain errors.
	
	In addition, in the results after executing the quantum circuit in Fig.\ref{figure7} and Fig.\ref{figure9}, we can see the probability that there are eight quantum states in both the plaintext and the ciphertext. According to Eq.\ref{eq5}, it can be known that after the quantum circuit is executed, the output quantum state $\left| {{\upsilon _2}} \right\rangle $ should only have four quantum states $\left\{ {010,011,100,101} \right\}$ at this time. This phenomenon is due to the fact that in the calculation process of the quantum system, the superposition state is generated to operate the entire system. As time progresses, it leads to decoherence, and the calculation information degrades over time. The phenomenon of bit flip occurs, and some quantum states are transformed into undesired quantum states. But for this kind of bit flip, the probability is the same, and the influence on the final experimental result is limited. Of course, the factors that affect the error are far more than these. As a complex quantum system, we cannot completely eliminate these factors.
	
	 In the second experiment, the same noise also exists. We simply verify the quantum circuits in Fig.\ref{figure13} and Fig.\ref{figure15} in a noise-free quantum simulation environment. The results only contain the expected quantum states $\left| {000} \right\rangle $ and $\left| {111} \right\rangle $. But the difference from the quantum circuit that executed $SwapTest$ in the first experiment is that the multi-value Grover quantum search executed in $GroverOpitm$ only needs to display the results we expect with a significantly high probability. The emergence of the quantum state has a limited impact on the experimental results, and the error can be reduced by performing the amplitude amplification process multiple times.

	\section{Conclusion and future works}
	\label{8}
	This paper proposes a quantum k-means algorithm based on trusted servers in quantum cloud computing. The core subroutines $SwapTest$ and $GroverOptim$ are encrypted by a quantum homomorphic encryption scheme and uploaded to the quantum cloud for calculation. Using a trusted server in the quantum cloud to simplify the tedious $T$-gate update steps. Reduce the overall computing efficiency and simplify the key update steps on the cloud. The client only needs to complete the encryption, decryption and update threshold operations, and other operations are handed over to the server to complete, which greatly reduces the workload of the client. 
	
	Compared with the classical k-means algorithm, the quantum k-means algorithm reduces the time complexity from $O\left( {Mnkt} \right)$ to $O\left( {M\log \left( n \right)\sqrt k t} \right)$. The time complexity of client encryption and decryption is polynomial, depending on the number of quantum states $m$ required. In addition, the quantum homomorphic encryption scheme based on trusted server is proposed in this paper, which ensures the security by using quantum one-time pad (QOTP) scheme. In addition, the quantum homomorphic encryption scheme based on trusted server is proposed in this paper, which ensures the security by using quantum one secret at a time. This scheme is not only suitable for quantum k-means algorithm, but also suitable for entrusted quantum computing, which is easy to describe, but complex computational process.
	
	We use IBM Qiskit to design and execute quantum circuits to verify the correctness of the scheme. The experimental results show that there is noise in the experimental environment. When the number of $H$ and $T$ gates in the circuit increases, the quantum state decoherently interacts with the environment, which leads to the degradation of computational information over time, making our experimental probability and accuracy both decline. 
	
	The core idea of this paper is to complete the subroutine of the quantum machine learning algorithm by the quantum cloud server, and to ensure the security through the quantum homomorphic encryption scheme. In addition, the idea can be used as a model for various commissioned quantum calculations that are easy to describe but have a long and complicated calculation process.
	
	Therefore, how to improve the scheme to reduce the complexity of the $T$-gate and reduce the number of times the system state is superimposed and de-superimposed and how to optimize the model to make the ciphertext computing operation based on quantum cloud computing meet more quantum machine learning algorithms is used will become our next research focus.



\end{document}